\documentclass[a4paper,11pt]{article}
\pdfoutput=1

\usepackage{jheppub}

\usepackage[T1]{fontenc}

\newcommand{\dd}{\mathrm{d}}

\title{Exact analytic rotating black-hole solutions with primary hair}

\author{Pedro G. S. Fernandes}
\affiliation{Institut f\"ur Theoretische Physik, Universit\"at Heidelberg, Philosophenweg 12 \& 16, 69120 Heidelberg, Germany}
\emailAdd{fernandes@thphys.uni-heidelberg.de}

\abstract{
Exact, analytic, asymptotically flat rotating black-hole solutions are exceedingly rare, with only a handful of examples known. Using a Kerr-Schild ansatz, we derive a multitude of exact, analytic, asymptotically flat rotating black-hole solutions within a broad class of Generalized Proca theories. These black holes differ significantly from Kerr black holes, as they possess primary hair and are non-circular, thus breaking a symmetry that vacuum black holes exhibit in General Relativity.
}

\begin{document} 
\maketitle
\flushbottom

\section{Introduction}

Schwarzschild obtained his celebrated spherically symmetric vacuum solution~\cite{1916SPAW.......189S} to the Einstein field equations less than a year after the introduction of General Relativity (GR), whereas the Kerr metric was discovered only some fifty years later~\cite{PhysRevLett.11.237}. The long delay is unsurprising: even after imposing stationarity and axial symmetry, the Einstein equations remain a highly non-linear system of partial differential equations.

Remarkably, however, the Kerr metric itself is exceedingly simple. It is the unique stationary, rotating, asymptotically flat vacuum black-hole solution of GR and is completely characterized by just two macroscopic parameters: its mass and angular momentum~\cite{PhysRevLett.26.331,PhysRevLett.34.905}. This striking simplicity underlies the \emph{no-hair conjecture}~\cite{Herdeiro:2015waa,Yazadjiev:2025ezx}, according to which gravitational collapse is expected to result in a spacetime geometry fully specified by these two parameters alone, with no additional independent degrees of freedom colloquially referred to as \emph{hair}. Moreover, the Kerr metric is also notably simple in that it satisfies a property known as \emph{circularity}~\cite{Babichev:2025szb,DelPorro:2025hse,Delaporte:2022acp}. Roughly speaking, circularity implies that a stationary, axisymmetric rotating black-hole metric can be expressed in coordinates where the metric contains a single off-diagonal component, and is invariant under the simultaneous reversal of the temporal and azimuthal coordinates. This symmetry is intimately connected, for example, with the separability of the geodesic equations~\cite{Babichev:2025szb}.

Beyond vacuum GR, only a few exact analytic four-dimensional rotating black-hole solutions are known that are asymptotically flat and regular on and outside the event horizon~\cite{KerrNewman,Sen:1992ua,Rasheed:1995zv,Larsen:1999pp} many of which are stealth\footnote{In this context, the term “stealth” refers to configurations in which the fields have non-trivial profiles but do not backreact on the metric, i.e., their combined stress-energy tensor vanishes.} Kerr metrics~\cite{Charmousis:2019vnf,Cisterna:2016nwq,Psaltis:2007cw,Xu:2026zgd}. In addition, new solutions related to known exact analytic metrics via field redefinitions can be obtained~\cite{Anson:2020trg,BenAchour:2020fgy,Filippini:2017kov,Tahara:2023pyg,Achour:2021pla,BenAchour:2020wiw,Babichev:2024eoh,BenAchour:2025lkx}.
Outside these very special cases, in order to study rotating solutions, perturbative approaches~\cite{Cano:2019ore,Candan:2025fbl,Charmousis:2021npl,Gammon:2022bfu,Maselli:2015yva,Hui:2021cpm,Kubiznak:2022vft,Gray:2021roq,Gray:2021toe,Maselli:2015tta,Ayzenberg:2014aka}, and numerical methods~\cite{Fernandes:2022gde,Dias:2015nua,Sullivan:2020zpf,Delgado:2020rev,Kleihaus:2011tg,Kleihaus:2015aje,Delsate:2018ome,Herdeiro:2020wei,Berti:2020kgk,Eichhorn:2025aja,Cunha:2019dwb,Staykov:2025lfh,Fernandes:2024ztk,Collodel:2019kkx,Garcia:2023ntf,Guo:2023mda,Xiong:2023bpl,Liu:2025bkz,Cheng:2025hdw,Herdeiro:2025blx,Liu:2025eve,Lam:2025elw,Lam:2025fzi,Herdeiro:2014goa,Herdeiro:2016tmi,Burrage:2023zvk,Fernandes:2025osu,Destounis:2025tjn,Fernandes:2025vxg} have been used, assuming circularity, since it reduces the number of independent components of a stationary and axially-symmetric to only four~\cite{Konoplya:2016jvv}.
The Newman-Janis algorithm~\cite{newmanNoteKerrSpinningParticle1965} is also frequently employed to generate rotating metrics from static ones; however, in the vast majority of cases, the resulting rotating metric does \emph{not} solve the field equations of the underlying theory and has pathologies~\cite{Hansen:2013owa}.

Most known exact solutions are circular and can either be expressed in Kerr-Schild form (or a generalized form thereof)~\cite{Debney:1969zz,Bini:2010hrs} or derived from a solution that admits such a form, through a solution generating method~\cite{Ayon-Beato:2015nvz,Ayon-Beato:2025ahb,Hassaine:2024mfs}. Kerr-Schild metrics can be written as
\begin{equation}
    g_{\mu \nu} = \eta_{\mu \nu} + 2 \Phi l_\mu l_\nu,
    \label{eq:kerrschild}
\end{equation}
where $\eta_{\mu \nu}$ is the Minkowski metric, $\Phi$ is a scalar function, and $l^\mu$ is the tangent vector to a shear-free and geodesic null congruence.
However, theories are generally not compatible with the Kerr-Schild ansatz, justifying the need for perturbative or numerical methods.
For example, the only theory of electrodynamics compatible with a Kerr-Schild ansatz is the usual Einstein-Maxwell theory~\cite{Kubiznak:2022vft}.

In this work, using a Kerr-Schild ansatz in the framework of Generalized Proca theory we discover not just one, but a multitude of exact, analytic, rotating black-hole solutions that violate circularity and possess \emph{primary hair} -- additional integration constants that influence the geometry and are not fixed by the mass or angular momentum of the black hole.

Our derivation of these exact solutions is strongly inspired by the standard Kerr-Schild construction of the Kerr-Newman metric, as well as by recent works in Proca theories of four-dimensional Gauss-Bonnet gravity~\cite{Charmousis:2025jpx,Eichhorn:2025pgy,Alkac:2025zzi,Liu:2025dqg,Konoplya:2025uiq,Alkac:2025jhx,Lutfuoglu:2025qkt,Konoplya:2025bte,Fernandes:2025mic}. The key observation is that, in the derivation of the Kerr-Newman solution, a drastic simplification of the field equations is achieved by aligning the electromagnetic field with the null vector $l^\mu$. Similarly, in the Proca theories considered in Refs.~\cite{Charmousis:2025jpx,Eichhorn:2025pgy}, the Proca field in the static, spherically symmetric case is precisely aligned along this direction, and the resulting spacetime geometry takes a Kerr-Schild form. These parallels motivated our search for exact rotating solutions exhibiting the same structural properties.

This paper is structured as follows. In Sec.~\ref{sec:Proca}, we present the framework of Generalized Proca theories and explain how a subclass of these theories, given in Eq.~\eqref{eq:GPrestricted}, is directly motivated by dimensional regularizations of the Lovelock invariants to four dimensions. Next, in Sec.~\ref{sec:eoms_rotating}, using a Kerr-Schild metric in ingoing Kerr coordinates~\eqref{eq:line-element}, we derive the independent equations of motion, given by Eqs.~\eqref{eq:MassFinal}-\eqref{eq:ProcaFinal}, and solve them explicitly in Sec.~\ref{sec:exact_solutions}. In Sec.~\ref{sec:disformal}, we introduce disformal transformations and show how they can be used to generate new rotating solutions in Generalized Proca theory. The breaking of circularity is discussed in Sec.~\ref{sec:circularity}. We conclude in Sec.~\ref{sec:conclusions} with a discussion of our results and possible directions for future work.

\section{Generalized Proca theory and four-dimensional Proca theories of Lovelock gravity}
\label{sec:Proca}
Generalized Proca~\cite{Heisenberg:2014rta} is a theory of gravity coupled to a vector field $A_\mu$ with second-order equations of motion that in addition to the dynamical degrees of freedom of the metric $g_{\mu \nu}$ propagates only two transverse modes and one longitudinal mode. It is described by\footnote{We work with units in which $G=1=c$.}
\begin{equation}
\label{Lgp}
    \mathcal{L}_{\text{gen.Proca}}=\sum_{n=2}^{6}\mathcal{L}_{n},
\end{equation}
\begin{align}
\label{eq:Lall}
    \mathcal{L}_2 &= G_2\!\left(A_\mu,\,F_{\mu\nu},\,\tilde F_{\mu\nu}\right),\notag\\[4pt]
    \mathcal{L}_3 &= G_3(X)\,\nabla_\mu A^\mu,\notag\\[4pt]
    \mathcal{L}_4 &= G_4(X)\,R
    + G_{4,X}\!\left[\,(\nabla_\mu A^\mu)^2-\nabla_\rho A_\sigma\,\nabla^\sigma A^\rho\right],\\[4pt]
    \mathcal{L}_5 &= G_5(X)\,G_{\mu\nu}\,\nabla^\mu A^\nu
    -\frac{1}{6}G_{5,X}\Big[(\nabla\!\cdot\!A)^3 + 2\,\nabla_\rho A_\sigma\,\nabla^\gamma A^\rho\,\nabla_\gamma A^\sigma
    - 3\,(\nabla\!\cdot\!A)\,\nabla_\rho A_\sigma\,\nabla^\sigma A^\rho\Big]\notag\\ 
    &\quad - g_5(X)\,\tilde F^{\alpha\mu}\,\tilde F^{\beta}{}_{\mu}\,\nabla_\alpha A_\beta,\notag\\[4pt]
    \mathcal{L}_6 &= G_6(X)\,\mathcal{P}^{\mu\nu\alpha\beta}\,\nabla_\mu A_\nu\,\nabla_\alpha A_\beta+\frac{G_{6,X}}{2}\,\tilde F^{\alpha\beta}\tilde F^{\mu\nu}\,\nabla_\alpha A_\mu\,\nabla_\beta A_\nu,\notag
\end{align}
where $X\equiv-A_{\mu}A^{\mu}/2$ is the norm of the Proca field, $F_{\mu\nu}\equiv\nabla_{\mu}A_{\nu}-\nabla_{\nu}A_{\mu}$ the field strength tensor with corresponding dual $\tilde{F}_{\mu\nu}$, and $\mathcal{P}^{\mu\nu\alpha\beta}$ the double-dual Riemann tensor.

Vector-tensor theories and the Generalized Proca class have been explored as modified gravity frameworks with rich phenomenology in both cosmology and black-hole physics~\cite{Kimura:2016rzw,Heisenberg:2016eld,DeFelice:2016uil,DeFelice:2016yws,BeltranJimenez:2016rff,Kase:2020yhw,Annulli:2019fzq,Heisenberg:2018acv,Heisenberg:2017hwb,Heisenberg:2017xda,deFelice:2017paw,Chagoya:2017fyl,Babichev:2017rti,Emami:2016ldl,BeltranJimenez:2016afo,Allys:2016kbq,Minamitsuji:2016ydr,Fernandes:2025lon,Tsujikawa:2025wca,DeFelice:2025ykh,Bertucci:2024qzt,Wall:2024lbd,Aoki:2023bmz,Ozsoy:2023gnl,Dong:2023xyb,Fell:2023mtf,Kase:2023kvq,Coates:2023dmz,Coates:2022nif,deRham:2022sdl,Coates:2022qia,Mou:2022hqb,Garcia-Saenz:2021uyv,Aoki:2021wew,Brihaye:2021qvc,Barton:2022rkj,Barton:2021wfj,Heisenberg:2020xak,Heisenberg:2020jtr,DeFelice:2020icf,Minamitsuji:2020pak,BeltranJimenez:2019wrd,ErrastiDiez:2019ttn,Colleaux:2025vtm,Colleaux:2024ndy,Colleaux:2023cqu,Oliveros:2019zkl,Rahman:2018fgy,Kase:2017egk,Heisenberg:2025roe,Chiang:2025gpa,Chen:2024hkm,Aoki:2023jvt,deRham:2020yet}.
More recently, motivated by Gauss-Bonnet gravity and its four-dimensional generalizations, which have attracted significant attention in recent years~\cite{Glavan:2019inb,Fernandes:2020nbq,Hennigar:2020lsl,Kobayashi:2020wqy,Lu:2020iav,Fernandes:2021dsb} (see Ref.~\cite{Fernandes:2022zrq} for a review), a Proca theory of four-dimensional Gauss-Bonnet gravity was constructed in Ref.~\cite{Charmousis:2025jpx}. This theory was obtained via a dimensional regularization of the Gauss-Bonnet invariant to four dimensions involving a Weyl vector, and belongs to a subset of the Generalized Proca class given by
\begin{equation}
    \mathcal{L} = G_2(X) + G_3(X) \nabla_\mu A^\mu + G_4(X) R + G_{4,X}\left[\,(\nabla_\mu A^\mu)^2-\nabla_\rho A_\sigma\,\nabla^\sigma A^\rho\right],
    \label{eq:GPrestricted}
\end{equation}
with functions defined by
\begin{equation}
    G_{2}=-24X^{2},\quad G_{3}=-16X, \quad G_{4}=-4X,
    \label{eq:GBfunctions}
\end{equation}
up to an overall common constant.

Following Refs.~\cite{Fernandes:2025fnz,Fernandes:2025eoc}, which constructed lower-dimensional scalar-tensor Lovelock theories of gravity, the Proca theory introduced in Ref.~\cite{Charmousis:2025jpx} was extended in Ref.~\cite{Fernandes:2025mic} by incorporating not only the Gauss-Bonnet invariant but all Lovelock invariants. The theory obtained by regularizing the $n^{\rm th}$ Lovelock invariant also falls within the restricted Generalized Proca class defined in Eq.~\eqref{eq:GPrestricted}, with the corresponding functions
\begin{equation}
    \begin{aligned}
    &G_2^{(n)}(X)=-6\cdot2^{n}(n-1)X^n,\\
    &G_3^{(n)}(X)=-2^{n+1}nX^{n-1},\\
    &G_4^{(n)}(X)=-2^{n-1}\frac{n}{2n-3}X^{n-1},
    \end{aligned}
    \label{eq:LovelockFunctions}
\end{equation}
where the Gauss-Bonnet case in Eq.~\eqref{eq:GBfunctions} is recovered for $n=2$.
These Proca-Lovelock theories have been explored in a variety of settings~\cite{Charmousis:2025jpx,Eichhorn:2025pgy,Alkac:2025zzi,Liu:2025dqg,Konoplya:2025uiq,Alkac:2025jhx,Lutfuoglu:2025qkt,Konoplya:2025bte,Fernandes:2025mic}. In particular, Refs.~\cite{Charmousis:2025jpx,Fernandes:2025mic} studied their static black-hole solutions, and found Kerr-Schild black holes with primary hair, and where the Proca field is aligned with the null direction $l^\mu$, such that $X=0$.

Motivated by these Proca theories arising from dimensional regularizations of Lovelock gravity, we now investigate black-hole solutions within the subset of the Generalized Proca class defined in Eq.~\eqref{eq:GPrestricted} with functions given by
\begin{equation}
    \begin{aligned}
        &G_2 = \mathcal{O}\left(X^2\right), \qquad G_3 = \alpha \,c_{3,1} X + \mathcal{O}\left(X^2\right),\\& G_4 = 1 + \alpha \, c_{4,1} X + \alpha^2\, c_{4,2} X^2 + \mathcal{O}\left(X^3\right),
    \end{aligned}
    \label{eq:ProcaFunctionsOnshell}
\end{equation}
which include the Lovelock-inspired class~\eqref{eq:LovelockFunctions} as a particular case. Terms proportional to higher powers of $X$ do \emph{not} contribute to the equations of motion, since in this work we consider only configurations with $X=0$ on-shell.

\section{Equations of motion for a rotating Kerr-Schild metric}
\label{sec:eoms_rotating}
The standard derivation of the Kerr-Newman metric relies on two key assumptions:
i) the metric is of Kerr-Schild type;
ii) the electromagnetic field is aligned with $l^\mu$.
With these assumptions, despite the increased complexity of having a non-zero stress-energy tensor, the Einstein-Maxwell field equations are easily integrable.
Motivated by the Kerr-Newman metric, and the static solutions of Refs.~\cite{Charmousis:2025jpx,Fernandes:2025mic}, we seek a rotating Kerr-Schild-type solution in which the Proca fields remain aligned with $l^\mu$ for the theory in Eq.~\eqref{eq:GPrestricted}\footnote{See Refs.~\cite{Babichev:2020qpr,Baake:2021jzv} for works that apply a similar idea in scalar-tensor theories to generate static and spherically symmetric regular black holes.}.

We consider the Kerr-Schild ansatz in ingoing Kerr coordinates
\begin{equation}
    \dd s^2 = -(1-2\Phi) (l_\mu \dd x^\mu)^2 + \Sigma (\dd \theta^2 + \sin^2 \theta \dd \varphi^2) + 2(\dd v + a \sin^2 \theta \dd \varphi)(\dd r + a \sin^2 \theta \dd \varphi),
    \label{eq:line-element}
\end{equation}
where we have defined $\Sigma = r^2 + a^2\cos^2\theta$, and $l_\mu \dd x^\mu = \dd v + a\sin^2 \theta \dd\varphi$, and $a$ is a parameter associated with rotation. In the following, we use the coordinate $\chi = \cos \theta$ for simplicity.

Following the standard derivation of the Kerr-Newman metric, we express the Kerr-Schild scalar function in terms of the generalized mass function of the spacetime
\begin{equation}
    \Phi = \frac{r M(r,\chi)}{\Sigma},
\end{equation}
and the Proca field is taken to be aligned with $l_\mu$
\begin{equation}
    A_\mu \dd x^\mu = \frac{\mathcal{A}(r,\chi)}{\Sigma} l_\mu \dd x^\mu,
    \label{eq:proca_profile}
\end{equation}
such that this configuration has $X=0$. When $M(r,\chi)=\mu\equiv \mathrm{cte}$, the Kerr geometry is recovered. As shown in Ref.~\cite{Ayon-Beato:2015nvz}, a metric of the form given in Eq.~\eqref{eq:line-element} is circular only if the mass function is independent of the angular coordinate, $M(r,\chi)\equiv M(r)$. Here we do not impose this restriction, which turns out to be crucial for deriving exact solutions.

The equations of motion of the Generalized Proca theory are obtained by varying the action with respect to the metric and the Proca field, yielding
\begin{equation}
    E_{\mu \nu} = 0,
    \label{eq:EinEqsCov}
\end{equation}
and
\begin{equation}
    \mathcal{V}^\mu = 0,
    \label{eq:ProcaEqsCov}
\end{equation}
respectively. The tensors $E_{\mu \nu}$ and $\mathcal{V}^\mu$ are given in Ref.~\cite{DeFelice:2016cri}. We have independently derived the equations of motion~\eqref{eq:EinEqsCov} and \eqref{eq:ProcaEqsCov} for the class~\eqref{eq:GPrestricted} using \texttt{xAct}~\cite{xact}.

A careful examination of the field equations of the theory~\eqref{eq:GPrestricted} reveals that there are only three independent equations which can chosen as
\begin{align}
    &\mathcal{E}_1 \equiv \partial_r \left( M - \frac{\mathcal{A}^2G_{4,X}}{2r \Sigma G_4} + \frac{(r^4+6a^2r^2\chi^2 - 3a^2 \chi^2)G_2}{12 r G_4} \right) = 0, \label{eq:FEQ1} \\
    &\mathcal{E}_2 \equiv \partial_\chi \left( M - \frac{\mathcal{A}^2G_{4,X}}{2r \Sigma G_4} \right) = 0, \label{eq:FEQ2} \\
    &\mathcal{E}_3 \equiv \partial_r \mathcal{A} - \frac{\Sigma^4 \left(G_{4,X}G_2 - G_4 G_{2,X} \right) + 2\mathcal{A}^2 \left( (3r^2+a^2\chi^2) G_{4,X}^2 + (3r^2-a^2\chi^2) G_4 G_{4,XX} \right)}{\Sigma^3 G_4 G_{3,X} + 4 \mathcal{A} r \Sigma (G_{4,X}^2+G_4 G_{4,XX})} = 0. \label{eq:FEQ3}
\end{align}
All other field equations can be written as linear combinations of these three equations and their derivatives, see Appendix \ref{app:B} for details.

Compatibility of $\mathcal{E}_1$ and $\mathcal{E}_2$ implies $G_2 = \mathcal{O}(X)$, and that
\begin{equation}
    M = \mu + \frac{\mathcal{A}^2G_{4,X}}{2r \Sigma G_4},
\end{equation}
where $\mu$ is an integration constant related to the mass of the solution.
Assuming the Proca functions can be expanded as in Eq.~\eqref{eq:ProcaFunctionsOnshell}, the field equations reduce to
\begin{align}
    &M = \mu + \alpha \, c_{4,1} \frac{\mathcal{A}^2}{2r\Sigma},
    \label{eq:MassFinal}\\
    &\partial_r \mathcal{A} = 2 \alpha \mathcal{A}^2\frac{3 r^2 \left(c_{4,1}^2+2 c_{4,2}\right)+a^2 \chi ^2 \left(c_{4,1}^2-2 c_{4,2}\right)}{ c_{3,1} \Sigma^3+4 \alpha  r \Sigma \left(c_{4,1}^2+2 c_{4,2}\right) \mathcal{A}}.
    \label{eq:ProcaFinal}
\end{align}
A few comments are in order. First, in the static limit ($a \to 0$), solving the differential equation~\eqref{eq:ProcaFinal} yields the solutions studied in Refs.~\cite{Charmousis:2025jpx,Fernandes:2025mic}. Second, the equation determining $\mathcal{A}$ is non-trivially decoupled from $M$. This implies that once a solution for $\mathcal{A}$ is obtained, a solution to the full system follows, thereby fully determining the metric.

\section{Exact analytic rotating solutions}
\label{sec:exact_solutions}

This section focuses on integrating Eq.~\eqref{eq:ProcaFinal} to obtain new rotating geometries via Eq.~\eqref{eq:MassFinal}. When $a \neq 0$ we were unable to find a general analytic solution to Eq.~\eqref{eq:ProcaFinal}. Instead, we proceed on a case-by-case basis by imposing specific relations among the coupling constants.
We found analytic solutions in particular cases, and in general, the solutions are non-circular. Moreover, they typically depend on a free \emph{integration function} -- a primary hair that depends on $\chi$ -- which arises from integrating Eq.~\eqref{eq:ProcaFinal}.

As a consequence of non-circularity, the radial coordinate location $H$ of the event horizon, is in general a function of the polar angle, obeying~\cite{Babichev:2025szb}
\begin{equation}
    H'(\theta)^2 + \Delta\rvert_{r=H(\theta)} = 0, \qquad \Delta = r^2+a^2-2r M(r,\theta),
    \label{eq:HorizonNonCircular}
\end{equation}
subject to the boundary conditions $H'(0)=H'(\pi)=0$. If the mass function is symmetric around the equator, then one of the boundary conditions at the poles can be substituted by the condition $H'(\pi/2)=0$.

For future convenience, we define the ADM mass as
\begin{equation}
    \mu_{\rm ADM} = \lim_{r\to \infty} M(r,\chi).
    \label{eq:adm_mass}
\end{equation}
Although a $\chi$ dependent ADM mass is an interesting possibility that warrants further investigation, in this work we impose the ADM mass does not depend on $\chi$.

\subsection{Solutions with $c_{4,2}=-c_{4,1}^2/2$}
\label{sec:c42_c41}
We start by considering the case $c_{4,2}=-c_{4,1}^2/2$. In this case, the Proca equation~\eqref{eq:ProcaFinal} simplifies to
\begin{equation}
    \partial_r \mathcal{A} = 4 \alpha \mathcal{A}^2\frac{a^2 \chi ^2 c_{4,1}^2}{ c_{3,1} \Sigma^3},
\end{equation}
which can be integrated to give the solution
\begin{equation}
    \mathcal{A} = q(\chi) \left[ 1 - \frac{\alpha q(\chi)c_{4,1}^2}{2c_{3,1} a^3 \chi^3} \left( \frac{a \chi r \left(3r^2 + 5 a^2 \chi^2  \right)}{\Sigma^2} + 3 \cot^{-1}\left( \frac{a \chi}{r}\right) - \frac{3}{2}\pi \right) \right]^{-1},
\end{equation}
where $q(\chi)$ is a free integration function.
The resulting mass function~\eqref{eq:MassFinal} is given by
\begin{equation}
    M = \mu + \frac{\alpha c_{4,1} q(\chi)^2}{2r\Sigma}\left[ 1 - \frac{\alpha q(\chi)c_{4,1}^2}{2c_{3,1} a^3 \chi^3} \left( \frac{a \chi r \left(3r^2 + 5 a^2 \chi^2  \right)}{\Sigma^2} + 3 \cot^{-1}\left( \frac{a \chi}{r}\right) - \frac{3}{2}\pi \right) \right]^{-2}.
    \label{eq:Solution1}
\end{equation}
Although not immediately obvious, this solution has a well-defined non-rotating limit\footnote{This needs not be the case, as $q(\chi)$ can be redefined in a $a$-dependent way.}, given by $\mathcal{A} = q(\chi)$. In this case, the geometry is spherically symmetric only when $q(\chi)\equiv q$, recovering the case studied in Ref.~\cite{Fernandes:2025mic}. The metric is asymptotically flat.
This mass function provides the first non-trivial example of a rotating solution in our framework.

\begin{figure}[]
	\centering
	\includegraphics[width=0.6\linewidth]{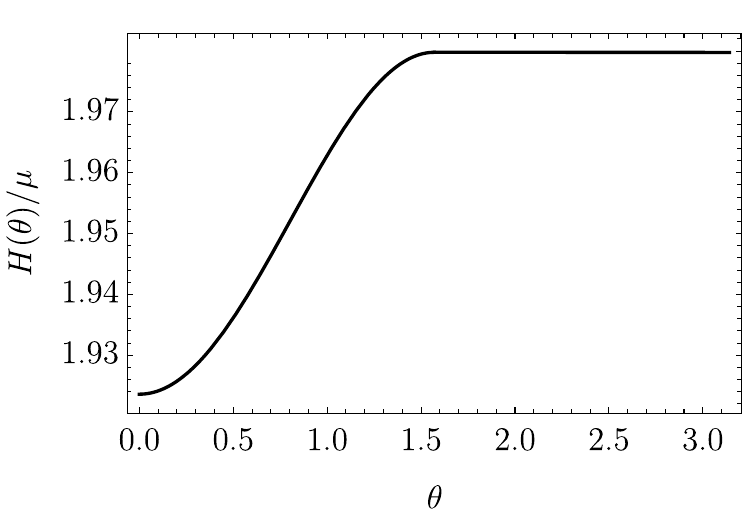}\hfill
    \caption{Radial coordinate location of the event horizon as a function of the polar angle for the metric described by Eq.~\eqref{eq:Solution1} with $q(\chi)=\mu \chi$, $c_{4,1}=-4$, $c_{3,1}=-16$, $a/\mu=0.2$, $\alpha/\mu^2=0.1$, and $\mu=1$.}
	\label{fig:HorizonNonSym}
\end{figure}

As an illustrative example showing that, in some cases, this solution represents a black hole, we solved Eq.~\eqref{eq:HorizonNonCircular} for the geometry~\eqref{eq:Solution1} using the parameter choices $q(\chi)=\mu \chi$, $c_{4,1}=-4$, $c_{3,1}=-16$, $a/\mu=0.2$, $\alpha/\mu^2=0.1$, and $\mu=1$. In Fig.~\ref{fig:HorizonNonSym}, we plot the horizon location as a function of the polar angle for this setup. We observe a non-trivial dependence on $\theta$ primarily in the range $\theta\in [0,\pi/2)$, while $H(\theta)$ remains \emph{approximately} constant, and close to the Kerr value, for $\theta\in [\pi/2,\pi]$. This behavior is shared by the mass function. We emphasize that this choice of parameters is intended merely to demonstrate that the solution in Eq.~\eqref{eq:Solution1} can describe a black hole, at least within a portion of the parameter space.

\subsection{Solutions with $c_{3,1}=0$}
\begin{figure}[]
	\centering
	\includegraphics[width=0.6\linewidth]{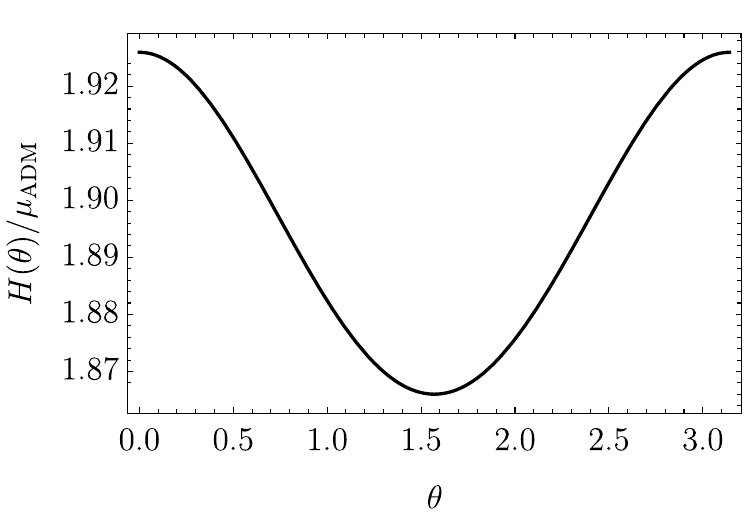}\hfill
    \caption{Radial coordinate location of the event horizon as a function of the polar angle for the metric described by Eq.~\eqref{eq:Solution_c31_0} with $\mu_{\rm ADM}=1$, $q/\mu_{\rm ADM}=1$, $c_{4,2}=-c_{4,1}^2$, $c_{4,1}=1$, $a/\mu_{\rm ADM}=0.5$, and $\alpha/\mu_{\rm ADM}^2=0.4$.}
	\label{fig:Horizonc31}
\end{figure}

When neglecting the Proca function $G_3$, by imposing $c_{3,1}=0$, Eq.~\eqref{eq:ProcaFinal} reduces to
\begin{equation}
    \partial_r \mathcal{A} = \mathcal{A} \frac{3 r^2 \left(c_{4,1}^2+2 c_{4,2}\right)+a^2 \chi ^2 \left(c_{4,1}^2-2 c_{4,2}\right)}{2 r \Sigma \left(c_{4,1}^2+2 c_{4,2}\right)},
\end{equation}
which has the solution\footnote{This parametrization for the integration function was chosen such that it has units of length.}
\begin{equation}
    \mathcal{A} = q(\chi)^{-\frac{1}{2}} r^{\frac{c_{4,1}^2-2c_{4,2}}{2\left(c_{4,1}^2+2c_{4,2}\right)}}\Sigma^{\frac{c_{4,1}^2+4c_{4,2}}{2\left(c_{4,1}^2+2c_{4,2}\right)}},
\end{equation}
resulting in a mass function given by
\begin{equation}
    M = \mu + \frac{\alpha c_{4,1}}{2} q(\chi)^{-1} \left(1 + \frac{a^2\chi^2}{r^2} \right)^{\frac{2c_{4,2}}{c_{4,1}^2+2c_{4,2}}}.
\end{equation}
Imposing the ADM mass~\eqref{eq:adm_mass} does not depend on $\chi$, we find that $q(\chi) \equiv q$, and
\begin{equation}
    \mu_{\rm ADM} = \mu + \frac{\alpha c_{4,1}}{2} q^{-1}.
\end{equation}
By writing the mass function in terms of the ADM mass, we get
\begin{equation}
    M = \mu_{\rm ADM} + \frac{\alpha\, c_{4,1}}{2}q^{-1} \left( \left(1 + \frac{a^2\chi^2}{r^2} \right)^{\frac{2c_{4,2}}{c_{4,1}^2+2c_{4,2}}}-1\right).
    \label{eq:Solution_c31_0}
\end{equation}
In the non-rotating limit we find that the geometry reduces to a stealth Schwarzschild metric. The solution~\eqref{eq:Solution_c31_0} also represents an asymptotically flat black hole, at least within a portion of the parameter space. As an example, we solved Eq.~\eqref{eq:HorizonNonCircular} choosing $\mu_{\rm ADM}=1$, $q/\mu_{\rm ADM}=1$, $c_{4,2}=-c_{4,1}^2$, $c_{4,1}=1$, $a/\mu_{\rm ADM}=0.5$, and $\alpha/\mu_{\rm ADM}^2=0.4$. The results are displayed in Fig.~\ref{fig:Horizonc31}, where the existence of a horizon is clearly demonstrated for this choice of parameters.

\subsection{Solutions with $c_{4,1}=0$: stealth Kerr}
From Eq.~\eqref{eq:MassFinal} we find that $M=\mu$, when $c_{4,1}=0$. Consequently, the geometry is simply a Kerr metric. However, the Proca field is non-trivial as the differential equation governing its profile becomes
\begin{equation}
    \partial_r \mathcal{A} = 4\alpha \mathcal{A}^2\frac{c_{4,2} \left( 3r^2 - a^2 \chi^2\right)}{c_{3,1}\Sigma^3 + 8 \alpha c_{4,2} r \Sigma \mathcal{A}},
\end{equation}
which has the solution
\begin{equation}
    \mathcal{A} = -\frac{c_{3,1} \Sigma^2}{8\alpha c_{4,2} r} \left( 1-\sqrt{1+\frac{128 c_{4,2}^2 \alpha q(\chi) r}{c_{3,1}^2 \Sigma^2}} \right).
    \label{eq:stealth_proca_c41=0}
\end{equation}
Consequently, the full solution is a stealth Kerr metric.

\section{Generating new rotating solutions via disformal transformations}
\label{sec:disformal}
Disformal transformations, reviewed for instance in Ref.~\cite{BenAchour:2024hbg}, have been employed in a variety of contexts, in particular as tools to generate new solutions from known ones, see Refs.~\cite{Anson:2020trg,BenAchour:2020fgy,Filippini:2017kov,Tahara:2023pyg,Achour:2021pla,BenAchour:2020wiw,Babichev:2024eoh,BenAchour:2025lkx}.

Although they are most commonly used in scalar-tensor theories, disformal transformations have also been generalized to the Proca case, for example in Ref.~\cite{Domenech:2018vqj}. These transformations can be expressed as
\begin{equation}
    g_{\mu \nu} \to \widetilde{g}_{\mu \nu} = g_{\mu \nu} + \beta A_\mu A_\nu,
    \label{eq:disformal}
\end{equation}
where $\beta$ is a constant and we ignored a conformal factor. Transformations of these type always map Generalized Proca theories to another set of Generalized Proca theories~\cite{Domenech:2018vqj}.
The transformation of the inverse metric follows
\begin{equation}
    \widetilde{g}^{\mu \nu} = g^{\mu \nu} - \frac{\beta}{1-2\beta X} A^\mu A^\nu,
\end{equation}
which is particularly simple when $X=0$, since on-shell it becomes
\begin{equation}
    \widetilde{g}^{\mu \nu} = g^{\mu \nu} - \beta A^\mu A^\nu.
\end{equation}
Consequently, for the class of solutions discussed here, a disformed metric remains a Kerr-Schild metric since
\begin{equation}
    \begin{aligned}
        &\widetilde{g}_{\mu \nu} = g_{\mu \nu} + \beta A_\mu A_\nu = \eta_{\mu \nu}+2\Phi l_\mu l_\nu + \beta A_\mu A_\nu \equiv \eta_{\mu \nu} + 2\widetilde{\Phi} l_\mu l_\nu,\\&
        \widetilde{g}^{\mu \nu} = g^{\mu \nu} - \beta A^\mu A^\nu = \eta^{\mu \nu}-2\Phi l^\mu l^\nu - \beta A^\mu A^\nu \equiv \eta^{\mu \nu} - 2\widetilde{\Phi} l_\mu l_\nu,
    \end{aligned}
\end{equation}
where we used that $A_\mu \propto l_\mu$. Starting from the line element in Eq.~\eqref{eq:line-element}, we find that a disformal metric is defined by the same line element but with
\begin{equation}
    \Phi\to \widetilde{\Phi} = \Phi + \frac{\beta \mathcal{A}(r,\chi)^2}{2\Sigma^2} \quad \Rightarrow \quad \widetilde{M}(r,\chi) = M(r,\chi) + \beta\frac{\mathcal{A}(r,\chi)^2}{2r\Sigma}.
\end{equation}
Upon using Eq.~\eqref{eq:MassFinal}, we find that the disformal metric is fully determined by the Proca field profile, since the mass function of the disformed metric reads
\begin{equation}
    \widetilde{M} = \mu + \left( \alpha\, c_{4,1} + \beta \right)\frac{\mathcal{A}^2}{2r\Sigma}.
\end{equation}
Therefore, the disformal transformation results in a geometry whose mass function has the same shape as before, the main difference being a shift $\alpha \, c_{4,1}\to \alpha \, c_{4,1}+\beta$ in the prefactor of the non-Kerr part of the mass function. There are, however, two exceptions where the qualitative behavior of the mass function changes drastically. First, any black-hole solution can always be disformed into a stealth Kerr metric by choosing $\beta = - \alpha \, c_{4,1}$, revealing the existence of a large class of Generalized Proca theories with stealth Kerr solutions. Second, when $c_{4,1}=0$, a stealth Kerr metric gets disformed into a non-trivial geometry.

\subsection{An example: a rotating Boulware-Deser solution}
Here we present an illustrative example of a solution generated via a disformal transformation. The Boulware-Deser black hole~\cite{PhysRevLett.55.2656} is an exact static and spherically symmetric solution of Einstein-Gauss-Bonnet gravity in higher dimensions. Following recent developments in four-dimensional Einstein-Gauss-Bonnet gravity~\cite{Glavan:2019inb,Fernandes:2020nbq,Hennigar:2020lsl,Kobayashi:2020wqy,Lu:2020iav,Fernandes:2021dsb}, the Boulware-Deser solution has been extended to four dimensions and is described by the line element in Eq.~\eqref{eq:line-element} with $a=0$ and
\begin{equation}
    M(r,\chi)\equiv M(r) = \mu - \frac{r^3}{8\alpha} \left( \sqrt{1+\frac{8\alpha \mu}{r^3}} -1\right)^2.
    \label{eq:boulware_deser}
\end{equation}
Note that the Proca theory of four-dimensional Einstein-Gauss-Bonnet gravity has a Boulware-Deser type metric with primary hair, found in Ref.~\cite{Charmousis:2025jpx}, that reduces to~\eqref{eq:boulware_deser} upon fixing the primary hair appropriately, in terms of the mass of the black hole.

Consider a disformal transformation of the stealth Kerr metric, whose Proca field profile is given in Eq.~\eqref{eq:stealth_proca_c41=0}, corresponding to the case $c_{4,1}=0$. The mass function of the disformal metric is given by
\begin{equation}
    \widetilde{M} = \mu + \beta \frac{c_{3,1}^2 \,\Sigma^3}{128c_{4,2}^2\alpha^2 r^3} \left( \sqrt{1+\frac{128c_{4,2}^2 \alpha q(\chi) r}{c_{3,1}^2\Sigma^2}} - 1 \right)^2.
\end{equation}
We find that the Boulware-Deser metric is recovered in the static limit, for instance by choosing $\beta=-\alpha$ and $c_{3,1}=4c_{4,2}$, and by fixing the integration constant in terms of the mass. By leaving the integration function free, the corresponding rotating metric is given by
\begin{equation}
    \widetilde{M} = \mu - \frac{\Sigma^3}{8\alpha r^3} \left( \sqrt{1+\frac{8 \alpha q(\chi) r}{\Sigma^2}} -1\right)^2,
    \label{eq:rotatingBD}
\end{equation}
representing a rotating Boulware-Deser type\footnote{In solutions of Einstein-Gauss-Bonnet gravity in vacuum, a linear combination of the Ricci and Gauss-Bonnet scalars vanishes. In this case, however, this is only true in the non-rotating limit and when $q(\chi)=\mu$.} black hole with primary hair.

\begin{figure}[]
	\centering
	\includegraphics[width=0.6\linewidth]{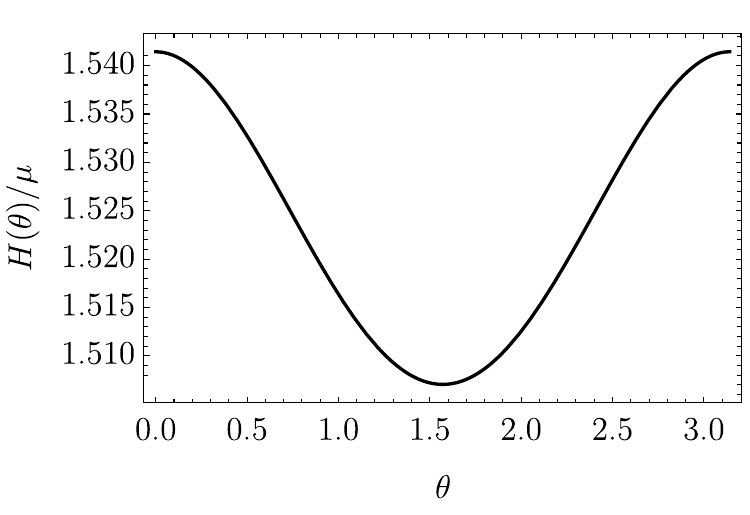}\hfill
    \caption{Radial coordinate location of the event horizon as a function of the polar angle for the metric described by Eq.~\eqref{eq:rotatingBD} with $q(\chi)=\mu$, $a/\mu=0.5$ and $\alpha/\mu^2=0.4$.}
	\label{fig:horizon}
\end{figure}

To verify that this geometry represents black holes, at least in some part of the respective parameter space, we have solved Eq.~\eqref{eq:HorizonNonCircular}. As an example, we present in Fig.~\ref{fig:horizon} the numerical solution to Eq.~\eqref{eq:HorizonNonCircular} for the rotating Boulware-Deser geometry~\eqref{eq:rotatingBD} with $q(\chi)=\mu$, and where we fixed $a/\mu=0.5$ and $\alpha/\mu^2=0.4$, demonstrating that the metric describes a black hole, at least in part of the parameter space.

\section{On the breaking of circularity}
\label{sec:circularity}

A stationary and axially-symmetric spacetime is said to be \emph{circular} if its Killing vectors are everywhere orthogonal to a family of codimension-two surfaces~\cite{Carter:1973rla,carterRepublicationBlackHole2009,carterRepublicationBlackHole2010}. Let $\xi^\mu$ and $\psi^\mu$ be the Killing vectors associated with stationarity and axial-symmetry, respectively, then a spacetime is circular if the following conditions hold
\begin{equation}
    \begin{aligned}
        &\xi^\mu R_{\mu[\nu}\xi_\rho \psi_{\sigma]} = 0,\\&
        \psi^\mu R_{\mu[\nu}\xi_\rho \psi_{\sigma]} = 0.
    \end{aligned}
    \label{eq:circularity}
\end{equation}

Circular, stationary, and axially symmetric black-hole spacetimes possess a number of properties that simplify their study. First, circularity ensures the existence of an adapted coordinate system (Boyer-Lindquist type) in which the metric has only a single off-diagonal component and is symmetric under the simultaneous reversal of the temporal and azimuthal coordinates. Such spacetimes can be fully described by just four functions of the radial and polar coordinates, which greatly facilitates the search for solutions using numerical methods~\cite{Xie:2021bur}. Additionally, circularity implies that key mechanical properties of the black hole, such as the angular velocity of the horizon and surface gravity, remain constant and that the event horizon is a Killing horizon. Circularity is also a necessary, though not sufficient, condition for the separability of geodesic motion~\cite{Babichev:2025szb}.

At the same time, the assumption of circularity may be overly restrictive, as it need not hold in all modified theories of gravity. While a theorem in Ref.~\cite{Xie:2021bur} asserts that black holes in effective field theories should be circular, this result relies on several assumptions, some of which can be easily circumvented in modified theories, see Ref.~\cite{Babichev:2025szb} for a detailed discussion on circumventing the assumptions of Ref.~\cite{Xie:2021bur}.
Non-circular black-hole geometries have been explored in Refs.~\cite{Anson:2020trg,Adam:2021vsk,Fernandes:2023vux,Grandclement:2023xwq,Delaporte:2022acp,Eichhorn:2021iwq,Eichhorn:2021etc,Chen:2023gwm}, and we refer the reader to the recent works~\cite{Babichev:2025szb,DelPorro:2025hse} for a comprehensive and modern treatment of non-circular black holes and their properties.

The line element in Eq.~\eqref{eq:line-element} represents a non-circular spacetime unless the mass function depends solely on the radial coordinate~\cite{Ayon-Beato:2015nvz}. As a result, with the exception of stealth Kerr solutions, the geometries obtained in this work are non-circular, exhibiting a non-trivial dependence on $\chi$. Our solutions evade the theorem of Ref.~\cite{Xie:2021bur} because the Proca field is "not invertible" in the sense defined in Ref.~\cite{Xie:2021bur} as it cannot be expressed as a linear combination of the two Killing vectors, and because some of the operators in the action~\eqref{eq:GPrestricted} with functions~\eqref{eq:ProcaFunctionsOnshell} have dimension not larger than four.

\section{Conclusions and future work}
\label{sec:conclusions}

Exact, analytic, asymptotically flat rotating black-hole solutions are exceedingly rare. Consequently, studies of rotating black holes beyond GR typically rely on perturbative or numerical approaches. In this work, we have constructed a large class of exact, analytic, asymptotically flat rotating black-hole solutions within a subset of Generalized Proca theories motivated by four-dimensional limits of Lovelock gravity. These solutions were obtained using a Kerr-Schild ansatz for the metric and a Proca field aligned with the null direction $l_\mu$, following closely the strategy used in the standard derivation of the Kerr-Newman metric.

The newly derived solutions are characterized by a single mass function and are fully determined, up to an integration function, once Eq.~\eqref{eq:ProcaFinal} is solved. While we did not find a general analytic solution to Eq.~\eqref{eq:ProcaFinal}, we obtained particular solutions by suitably choosing the coupling constants. Additionally, we generated new solutions via disformal transformations, including, notably, a rotating Boulware-Deser type metric.

Our work opens several new avenues for research. First, it shows that exact, analytic, asymptotically flat black-hole solutions can be obtained in a far broader class of modified theories than previously recognized, by employing a Kerr-Schild ansatz. A natural question is which other theories, or extensions of the class studied here, admit exact solutions. For instance, we did not include terms in $\mathcal{L}_5$ of the Generalized Proca Lagrangian in Eq.~\eqref{eq:Lall}, as they are not directly motivated by dimensional reductions of Lovelock gravity. Our preliminary analysis in the spherically symmetric case shows that the field equations can be integrated in some instances, suggesting that a detailed study of the rotating case would be an interesting direction for future work.

The integration function appearing in our exact solutions leaves residual freedom in fully specifying the geometry. As a direction for future research, it would be interesting to determine whether this integration function is a generic feature of non-circular black-hole solutions, potentially fixed by the dynamical evolution of the system as it relaxes toward an equilibrium rotating black hole.

No canonical kinetic term for the Proca field was included, as it is not motivated by the dimensional reduction of the Lovelock terms. We did, however, consider the effect of adding a term $\sim F_{\mu \nu} F^{\mu \nu}$ to the theories under study and found that it generally spoils integrability, even in the spherically symmetric case. Interestingly, terms containing the canonical kinetic term are naturally generated under disformal transformations~\eqref{eq:disformal} (see Ref.~\cite{Domenech:2018vqj}). A detailed investigation of the theories obtained via disformal transformations of the class considered here is an interesting direction for future research.

Although a general analytic solution to Eq.~\eqref{eq:ProcaFinal} could not be obtained, it should be possible to solve it numerically. Doing so would allow one to construct the rotating solution of the four-dimensional Einstein-Gauss-Bonnet theory of Ref.~\cite{Charmousis:2025jpx}. It should be noted, however, that the exact solutions presented in Sec.~\ref{sec:c42_c41} correspond to a specific reduction of the quadratic and cubic Lovelock invariants, with couplings satisfying particular relations as detailed in Ref.~\cite{Fernandes:2025mic}. Hence, these solutions can be viewed as concrete examples of rotating four-dimensional Lovelock black holes.

Due to the non-circularity of these spacetimes, geodesic motion is  non-integrable, which can give rise to intriguing phenomena such as chaotic orbital dynamics~\cite{Chen:2023gwm} and cusps in black-hole shadows~\cite{Eichhorn:2021etc,Eichhorn:2021iwq}. These effects could have important implications for both gravitational-wave and very-long-baseline interferometry observations. It would be particularly interesting to explore them in greater detail for the solutions presented here, as these black holes possess primary hair that can significantly deform the geometry away from Kerr, even when the length scale associated with the coupling constant $\alpha$ is small.
In addition, the mechanics of these black holes~\cite{DelPorro:2025hse} and their singularity structure constitute promising directions for future research. A preliminary analysis suggests that, in many cases, the singularity structure resembles that of a Kerr black hole.

\acknowledgments
The author thanks Astrid Eichhorn, Christos Charmousis and Mokhtar Hassaine for interesting discussions and comments on a first version of the manuscript.
This work is funded by the Deutsche Forschungsgemeinschaft (DFG, German Research Foundation) under Germany’s Excellence Strategy EXC 2181/1 - 390900948 (the Heidelberg STRUCTURES Excellence Cluster).

\appendix
\section{Linearly independent field equations}
\label{app:B}

The Einstein equations for the Generalized Proca theory~\eqref{eq:GPrestricted}, when considering a Kerr-Schild ansatz in ingoing Kerr coordinates~\eqref{eq:line-element}, and a null Proca field with $X=0$, can be written in matrix form as
\begin{equation}
     \mathbf{E} = \mathbf{k_1} \mathcal{E}_1 + \mathbf{k_2} \partial_r \mathcal{E}_1 + \mathbf{k_3} \mathcal{E}_2 + \mathbf{k_4} \partial_\chi \mathcal{E}_2 + \mathbf{k_5} \partial_r \mathcal{E}_2 + \mathbf{k_6} \mathcal{E}_3,
     \label{eq:FieldEqsMatrixForm}
\end{equation}
where $\mathbf{E}$ is the matrix representing $E_{\mu}^{\phantom{\mu}\nu}$ in this coordinate system, and the $\mathbf{k_i}$ are $4\times 4$ matrices given by
\begin{equation}
\mathbf{k_1}=\left(
\begin{array}{cccc}
 -\frac{2 G_{4} \left(a^4 \chi ^2 \left(\chi ^2-1\right)+a^2 r^2+r^4\right)}{\Sigma^3} & -\frac{2 r^2 G_{4,X} \mathcal{A}^2}{\Sigma^4} & 0 & \frac{2 a G_{4} \left(r^2-a^2 \chi ^2\right)}{\Sigma^3} \\
 0 & -\frac{2 r^2 G_{4}}{\Sigma^2} & 0 & 0 \\
 0 & 0 & -\frac{2 a^2 \chi ^2 G_{4}}{\Sigma^2} & 0 \\
 \frac{2 a \left(1-\chi ^2\right) \left(a^2+r^2\right) G_{4} \left(a^2 \chi ^2-r^2\right)}{\Sigma^3} & \frac{2 a r^2 \left(\chi ^2-1\right) G_{4,X} \mathcal{A}^2}{\Sigma^4} & 0 & -\frac{2 a^2 G_{4} \left(a^2 \chi ^2+r^2 \left(2 \chi ^2-1\right)\right)}{\Sigma^3} \\
\end{array}
\right),
\end{equation}

\begin{equation}
\mathbf{k_2}=\left(
\begin{array}{cccc}
 -\frac{a^2 r \left(\chi ^2-1\right) G_4}{\Sigma^2} & 0 & 0 & -\frac{a r G_4}{\Sigma^2} \\
 0 & 0 & 0 & 0 \\
 0 & 0 & -\frac{r G_4}{a^2 \chi ^2+r^2} & 0 \\
 -\frac{a r \left(\chi ^2-1\right) \left(a^2+r^2\right) G_4}{\Sigma^2} & 0 & 0 & -\frac{r \left(a^2+r^2\right) G_4}{\Sigma^2} \\
\end{array}
\right),
\end{equation}

\begin{equation}
\mathbf{k_3}=\left(
\begin{array}{cccc}
 \frac{2 a^2 r \chi  \left(\chi ^2-1\right) G_4}{\Sigma^3} & \frac{2 r \chi  \left(a^2+r^2\right) G_4}{\Sigma^3} & \frac{\left(\chi ^2-1\right) G_4 \left(r^2-a^2 \chi ^2\right)}{\Sigma^3} & \frac{2 a r \chi  G_4}{\Sigma^3} \\
 0 & 0 & 0 & 0 \\
 0 & \frac{G_4 \left(a^2 \chi ^2-r^2\right)}{\Sigma^2} & 0 & 0 \\
 -\frac{2 a^3 r \chi  \left(\chi ^2-1\right)^2 G_4}{\Sigma^3} & -\frac{2 a r \chi  \left(\chi ^2-1\right) G_4 \left(a^2 \left(\chi ^2+1\right)+2 r^2\right)}{\Sigma^3} & \frac{a \left(\chi ^2-1\right)^2 G_4 \left(a^2 \chi ^2-r^2\right)}{\Sigma^3} & -\frac{2 a^2 r \chi  \left(\chi ^2-1\right) G_4}{\Sigma^3} \\
\end{array}
\right),
\end{equation}

\begin{equation}
\mathbf{k_4}=\left(
\begin{array}{cccc}
 0 & \frac{r \left(\chi ^2-1\right) G_4}{\Sigma^2} & 0 & 0 \\
 0 & 0 & 0 & 0 \\
 0 & 0 & 0 & 0 \\
 0 & -\frac{a r \left(\chi ^2-1\right)^2 G_4}{\Sigma^2} & 0 & 0 \\
\end{array}
\right),
\end{equation}

\begin{equation}
\mathbf{k_5}=\left(
\begin{array}{cccc}
 0 & 0 & -\frac{r \left(\chi ^2-1\right) G_4}{\Sigma^2} & 0 \\
 0 & 0 & 0 & 0 \\
 0 & \frac{r G_4}{a^2 \chi ^2+r^2} & 0 & 0 \\
 0 & 0 & \frac{a r \left(\chi ^2-1\right)^2 G_4}{\Sigma^2} & 0 \\
\end{array}
\right),
\end{equation}

\begin{equation}
\mathbf{k_6}=\left(
\begin{array}{cccc}
 0 & -\frac{\mathcal{A}^2 \left(G_4 \left(4 r G_{4,XX} \mathcal{A}+\Sigma^2 G_{3,X}\right)+4 r G_{4,X}^2 \mathcal{A}\right)}{2 G_4 \Sigma^5} & 0 & 0 \\
 0 & 0 & 0 & 0 \\
 0 & 0 & 0 & 0 \\
 0 & \frac{a \left(\chi ^2-1\right) \mathcal{A}^2 \left(G_4 \left(4 r G_{4,XX} \mathcal{A}+\Sigma^2 G_{3,X}\right)+4 r G_{4,X}^2 \mathcal{A}\right)}{2 G_4 \Sigma^5} & 0 & 0 \\
\end{array}
\right).
\end{equation}
Also, the only non-trivial component of the Proca field equations~\eqref{eq:ProcaEqsCov} can be expressed as
\begin{equation}
    \mathcal{V}^{r} = -\frac{4r^2\mathcal{A} G_{4,X}}{\Sigma^3} \mathcal{E}_1 - \frac{\mathcal{A} \left( 4r \mathcal{A} G_{4,X}^2 + G_4 \left( \Sigma^2 G_{3,X} + 4 r \mathcal{A} G_{4,XX} \right) \right)}{\Sigma^4 G_4} \mathcal{E}_3,
\end{equation}
demonstrating that Eqs.~\eqref{eq:FEQ1}, \eqref{eq:FEQ2}, and \eqref{eq:FEQ3} provide a complete set of linearly independent field equations.

\bibliographystyle{JHEP}

\bibliography{Bibliography.bib}

\end{document}